\documentclass[twoside]{aiml16}

\usepackage{aiml16macro}

\def\lastname{Woltzenlogel Paleo}

\usepackage{xspace}

\usepackage{amsmath}
\usepackage{amssymb}

\newcommand{\R}{\mathbb{R}} 

\newcommand{\Q}{\mathcal{P}} 
\newcommand{\probab}{P} 

\newcommand{\Lang}{\mathcal{L}} 

\newcommand{\llist}{\texttt{list}} 
\newcommand{\nil}{\texttt{nil}} 
\newcommand{\cons}{::}

\newcommand{\imp}{\rightarrow}
\newcommand{\biimp}{\leftrightarrow}
\newcommand{\al}{\forall}
\newcommand{\ex}{\exists}

\newcommand{\ldia}{\Diamond}
\newcommand{\lbox}{\Box}

\newcommand{\In}{\texttt{in}}

\newcommand{\True}{\mathbf{T}}
\newcommand{\False}{\mathbf{F}}

\begin{document}

\begin{frontmatter}
  \title{An Expressive Probabilistic Temporal Logic}
  \author{Bruno Woltzenlogel Paleo}
  
  \begin{abstract}
  This paper argues that a combined treatment of probabilities, time and actions is essential for an appropriate logical account of the notion of probability; and, based on this intuition, describes an expressive probabilistic temporal logic for reasoning about actions with uncertain outcomes. The logic is \emph{modal} and \emph{higher-order}: modalities annotated by actions are used to express possibility and necessity of propositions in the next states resulting from the actions, and a higher-order function is needed to express the probability operator. The proposed logic is shown to be an adequate extension of classical mathematical probability theory, and its expressiveness is illustrated through the formalization of the Monty Hall problem.
  \end{abstract}

\end{frontmatter}

\newcommand{\PTL}{\textbf{PTL}\xspace}

\section{Introduction}

\noindent 
In order to reason about probabilistic knowledge, we must reason about time and actions as well. When we say, for example, that ``the probability of `heads' after a coin toss is 50\% and that of `tails' is 50\%'', we implicitly assume that there is an action (in this example, tossing a coin) which can bring the world to different states in the next moment in time. The uncertainty lies in the state transition: the world may end up in a state where the coin shows heads or in a state where it shows tails. 

Despite the evident dependence of our informal notion of probability on the notions of action and time, the formal mathematical languages that we use to talk about probabilities rarely support mentioning action and time explicitly. Kolmogorov's probability theory, for example, merely defines probability as the measure function in a measure space with total measure 1 \cite{Kolmogorov}. The task of modeling time-dependent actions and their possible outcomes in terms of events in a probabilistic space remains informal. While this informality is not problematic in the simplest situations (e.g. when we are interested in the possible outcomes of a single action, or when multiple actions are independent of each other), slightly more complex situations may already lead to confusion and difficulty. A famous example is the Monty Hall problem \cite{MontyHall}.

Another inconvenience of dealing with probabilities just in terms of a measure space is that its set-theoretic language (where events are represented as subsets of the sample space) is rather limited. There are obvious parallels between, for instance, set intersection and conjunction or set union and disjunction, which allow us to represent \emph{propositional} probabilistic knowledge easily (e.g. the event of a randomly picked coin showing heads \emph{and} the same coin being made of silver can be represented as the \emph{intersection} of the event of showing heads with the event of being made of silver). However, it is not clear how this analogy could be extended to more expressive logics with quantifiers. 

The main contribution of this paper, addressing the above mentioned issues, is the development of the syntax (in Section \ref{sec:Syntax}) and the semantics (in Section \ref{sec:Semantics}) of an expressive probabilistic temporal logic (\PTL) for reasoning about actions with uncertain outcomes. \PTL is an adequate extension of classical probability theory (as demonstrated in Section \ref{sec:Adequacy}), and its greater expressiveness allows us to reason explicitly about event independence (as discussed in Section \ref{sec:Independence}) and to avoid typical ambiguities of natural language discourse about probabilities (as shown in Section \ref{sec:Disambiguation}). This capacity of \PTL to avoid ambiguities related to outcomes and events is one of its main conceptual novelties in comparison to related work (cf. Section \ref{sec:RelatedWork}). \PTL's convenience and expressive power are illustrated through the formalization of the Monty Hall problem (in Section \ref{sec:MontyHall}).


\section{Syntax}
\label{sec:Syntax}

The aim of \PTL's language is to be sufficiently expressive to capture typical probabilistic statements, conveniently similar to natural language, and yet more precise than natural language in cases when the latter is ambiguous. Intuitively, probability is an inherently higher-order function, since it takes a proposition (representing an event) as an argument. Therefore, if a probabilistic logical language is to include a probability operator in the syntactic level, it is only natural that it should be a higher-order language.
Furthermore, because thinking probabilistically involves numerical computation and reasoning about states and actions, it is convenient to have a typed language, with distinct basic types for numbers, states and actions. The types used here are mostly the well-known simple types, but a list type constructor is included as well, in order to allow the representation of temporal sequences of actions and propositions.

\begin{definition}
\emph{Types} are freely generated from the set of basic types $\{ \beta, \iota, \eta, \mu \}$, the right-associative function type constructor $\imp$ and the list type constructor $\llist$. $\mu$ is the type for \emph{states}, $\beta$ is the type for \emph{booleans}, $\iota$ is the type for \emph{objects} and $\eta$ is the type for \emph{real numbers}. The set of all types is denoted $T$. The type of (local) \emph{propositions} $o$ is defined to be an abbreviation for $\mu \imp \beta$ and the type of actions $\alpha$ is defined to be an abbreviation for $\mu \imp \llist[\mu]$. 
\end{definition}

\begin{remark}
The definition of $o$ ensures that the truth of a proposition depends on states. The definition of $\alpha$ follows the intuition that an action can be seen as a function that maps a state to a list of possible next states.
\end{remark}

As shown in Definition \ref{def:Symbols}, \PTL contains, besides the usual logical symbols, also symbols for arithmetical functions and relations, the hybrid logic symbols for explicitly referring to states, list constructors and functions, and a probability operator. As modal operators ($\lbox$ and $\ldia$) implicitly bind states, they have a more fundamental role, which reminds that of the $\lambda$ binder. Therefore, they are treated separately in Definition \ref{def:Expressions}.

\begin{definition}
\label{def:Symbols}
For every type $\tau$, $S_{\tau}$ is a countably infinite set of uninterpreted symbols of type $\tau$. The set of arithmetic function symbols $S_{AF}$ is the set $\{ 0_{\eta}, 1_{\eta}, +_{\eta \imp \eta \imp \eta}, *_{\eta \imp \eta \imp \eta} \}$. The set of arithmetic relation symbols $S_{AR}$ is $\{ =_{\eta \imp \eta \imp o}, <_{\eta \imp \eta \imp o} \}$. The set of propositional logical symbols $S_L$ is $\{ \top_o, \bot_o, \vee_{o \imp o \imp o}, \wedge_{o \imp o \imp o}, \imp_{o \imp o \imp o}, \biimp_{o \imp o \imp o}, \neg_{o \imp o} \}$. The set of hybrid logical symbols $S_H$ is $\{ @_{\mu \imp o \imp o}, \In_{\mu \imp o} \}$. The set of quantifiers $S_Q$ is $\bigcup_{\tau \in T} \{\al_{(\tau \imp o) \imp o}, \ex_{(\tau \imp o) \imp o}, =_{o \imp o \imp o}  \}$. The symbol $\nil^\tau$ has type $\llist[\tau]$, $\cons^\tau$ has type $\tau \imp \llist[\tau] \imp \llist[\tau]$, $\in^{\tau}$ has type $\tau \imp \llist[\tau] \imp o$ and the length operator $|.|^{\tau}$ has type $\llist[\tau] \imp \mu$. The probability operator $\Q$ has type $\llist[\alpha] \imp o \imp \eta$. The set of all symbols $S$ is defined as $S_\tau \cup S_{AF} \cup S_{AR} \cup S_L \cup S_H \cup S_Q \cup \bigcup_{\tau \in T} \{\nil^{\tau}, \cons^{\tau}, |.|^{\tau}, \in^{\tau}\} \cup \{ \Q \}$. 
\end{definition}

Expressions are constructed as in the lambda calculus, using the symbols from $S$, application, abstraction and modalities.

\begin{definition}
\label{def:Expressions}
\emph{Expressions} are constructed according to the following rules:
\begin{itemize}
\item if $s_{\tau} \in S$, then $s_{\tau}$ is an expression of type $\tau$.
\item if $t_1$ is an expression of type ${\tau \imp \tau'}$ and $t_2$ is an expression of type $\tau$, then $(t_1~t_2)$ is an expression of type $\tau'$.
\item if $x_{\tau} \in S_{\tau}$ and $t$ is an expression of type $\tau'$, then $\lambda x_{\tau}. t$ is an expression of type $\tau \imp \tau'$.
\item if $\varphi$ is an expression of type $o$, $p$ is an expression of type $\eta$ and $a$ is an expression of type $\alpha$, then $\ldia^p_a \varphi$ are $\lbox_a \varphi$ expressions of type $o$.
\end{itemize}
\emph{Formulas} are expressions of type $o$. \emph{Actions} are expressions of type $\alpha$. The set of expressions of type $\tau$ is denoted $E_{\tau}$. $\Lang = \bigcup_{\tau \in T} E_{\tau}$.
\end{definition}

\begin{remark}
Types are omitted when they can be inferred from the context.
The usual parenthesis conventions are followed. Numerals are occasionally written in decimal notation. Infix notation is employed as usual for logical connectives, arithmetical functions and relations and the list constructor $\cons$. Binding notation is used for quantifiers.
Additionally, the following notation conventions and abbreviations are used:
\begin{itemize}
\item $\ldia_a \varphi \equiv \ex x_{\eta}. \ldia^{x}_a \varphi$

\item $\Q_l(\varphi) \equiv ((\Q~l)~\varphi)$

\item $\al x:G.~H(x) \equiv \al x.~G(x) \imp H(x)$

\item $\ex x:G.~H(x) \equiv \ex x.~G(x) \wedge H(x)$

\item $\al x_{\tau} \in \ell_{\llist[\tau]}.~H(x) \equiv \al x.~ (x \in \ell) \imp H(x)$

\item $\ex x_{\tau} \in \ell_{\llist[\tau]}.~H(x) \equiv \ex x.~ (x \in \ell) \wedge H(x)$

\item $\Q_{a\cons l}(\varphi \cons L) \equiv \Q_{a\cons l}(\varphi) \wedge \Q_{l}(L)$ (with $\Q_{\nil}(\nil) \equiv \top$)
\end{itemize}
\end{remark}

Probabilities appear in the logical language in two ways: firstly, as annotations on the diamond modal operator, in order to indicate how probable the corresponding state transition is; and secondly, through the higher-order probability function $\Q$, which takes a list of actions and a proposition as arguments and returns the probability that the proposition will hold after the execution of the listed actions.

\begin{example}
The following are some simple examples of probabilistic statements and their corresponding formalizations in \PTL:
\begin{enumerate}
\item Tossing a coin has a transition with probability $0.5$ to a state where the coin shows heads:
$
\al x:\mathit{Coin}. \ldia^{0.5}_{\mathit{toss}(x)} \mathit{heads}(x)
$
\item The probability of a coin showing heads after it is tossed is $0.5$: 
\[
\al x:\mathit{Coin}. \Q_{\mathit{toss}(x)\cons\nil}(\mathit{heads}(x)) = 0.5
\]
\item The probability of a coin showing heads twice after it is tossed twice is less than $0.5$: 
$
\al x:\mathit{Coin}. \Q_{\mathit{t}(x)\cons\mathit{t}(x)\cons\nil}(\mathit{h}(x)\cons\mathit{h}(x)\cons\nil) < 0.5
$ , 
where $t = \mathit{toss}$ and $h = \mathit{heads}$.
\item After a coin is tossed it is necessarily either heads or tails:
\[
\al x:\mathit{Coin}. \lbox_{\mathit{toss}(x)} (\mathit{heads}(x) \vee \mathit{tails}(x))
\]
\item After a coin is tossed it is possibly tails:
$
\al x:\mathit{Coin}. \ldia_{\mathit{toss}(x)} (\mathit{tails}(x))
$
\end{enumerate}
\end{example}

\section{Semantics}
\label{sec:Semantics}

For each type $\tau$, we need a domain $D_{\tau}$ of elements on which expressions of type $\tau$ are interpreted. For numerical expressions, we assume the domain to be a real closed field. For booleans, we assume the set with the usual two truth values. For function types, we require \emph{all} functions to be present in the type's domain. This effectively results in a \emph{standard} higher-order semantics. For \emph{Henkin} semantics, it would suffice to drop this last condition.

\begin{definition}
A \emph{domain} $D_{\tau}$ for a type $\tau$ is a non-empty set such that $D_{\tau' \imp \tau}$ is the set of all functions from $D_{\tau'}$ to $D_{\tau}$ (for every $\tau'$ and $\tau$), $D_o = \{ \True, \False \}$, $D_{\eta} = \R$ and $D_{\llist[\tau]}$ is the set of all lists of elements from $D_{\tau}$.
\end{definition}

As in the most common modal logics \cite{Blackburn}, we use \emph{frames} as the foundation for the modal aspects of the semantics. A frame is essentially a set of states and a relation for the transitions between states. What is different here is that transitions are labeled by actions and by probabilities, and the transition relation and actions must be mutually consistent.

\begin{definition}
\label{def:Frame}
A \emph{probabilistic labeled frame} is a triple $(W,R,\probab)$ such that $W$ is a non-empty set of \emph{states}, $R \subseteq W \times W \times D_{\alpha}$ satisfying the condition that if $(w, w', \ell) \in R$ then $(w, w'', \ell) \in R$ for every $w'' \in \ell(w)$, and $\probab: R \imp [0,1]$ is a probability function satisfying the condition that
for all $w \in W$ and for all $\ell \in D_{\alpha}$ such that there exists $w' \in W$ with $(w, w', \ell) \in R$, 
$$
\sum_{w' ~ | ~ (w, w', \ell) \in R } \probab((w,w',\ell)) = 1 
$$
\end{definition}

\begin{remark}
The relation $R$ in definition \ref{def:Frame} may be cyclic. This is convenient, for instance, when specifying Markov chains. 
\end{remark}

A model extends a frame with an interpretation function that assigns denotations to expressions. The denotation of an expression may generally vary with the state. In such cases, we say that the interpretation is \emph{flexible}; otherwise, it is \emph{rigid} \cite{Fitting}. In the examples considered in this paper, boolean expressions and probabilistic expressions are always flexibly interpreted, whereas other expressions are always rigidly interpreted.

\begin{definition}
\label{def:Model}
A \emph{model} is a tuple $(W, R, \probab, \{ D_{\tau} \}_{\tau \in T}, I)$ where $(W, R, \probab)$ is a probabilistic labelled frame, $\{ D_{\tau} \}_{\tau \in T}$ is a domain, $W = D_{\mu}$ and $I$ is an interpretation function that maps states and expressions of any type $\tau$ to elements in $D_{\tau}$. It is assumed that any interpretation $I$ maps arithmetic symbols, list constructors and functions, and logical constants to their usual fixed denotations. Therefore (as usual, non-exhaustively):
\begin{small} 
\begin{itemize}
\item $I_w(A \wedge B) = \True$ iff $I_w(A) = \True$ and $I_w(B) = \True$
\item $I_w(A \vee B) = \True$ iff $I_w(A) = \True$ or $I_w(B) = \True$
\item $I_w(A \imp B) = \True$ iff $I_w(A) = \False$ or $I_w(B) = \True$
\item $I_w(\neg A) = \True$ iff $I_w(A) = \False$
\item $I_w(\al x_{\tau}. \varphi) = \True$ iff $I_w[x \mapsto e](\varphi) = \True$ for every $e \in D_{\tau}$
\item $I_w(\ex x_{\tau}. \varphi) = \True$ iff $I_w[x \mapsto e](\varphi) = \True$ for some $e \in D_{\tau}$
\item $I_w( (t_1~t_2) ) = (I_w(t_1)~I_w(t_2))$
\item $I_w(\lambda x_{\tau}. t)$ is the function taking an element $e \in D_{\tau}$ and returning $I_w[x \mapsto e](t)$.
\item $I_w(\In(s)) = \True$ iff $w = I_w(s)$

\item $I_w(@_s \varphi) = \True$ iff $I_{I_w(s)}(\varphi) = \True$
\end{itemize}
\end{small}
where $I_w[x \mapsto e](x) = e$ and $I_w[x \mapsto e](t) = I[t]$ for any $t$ distinct from $x$.

\medskip

\noindent
Furthermore, and most importantly, the interpretations of expressions formed with modal and probabilistic operators are defined as follows:
\begin{small}
\begin{itemize}

\item $I_w(\lbox_a \varphi) = \True$ iff $I_{w'}(\varphi) = \True$ \\ for every $w'$ such that $(w, w', I_w(a)) \in R$ 

\item $I_w(\ldia_a^p \varphi) = \True$ iff $\probab((w,w',I_w(a))) = I_w(p)$ and $I_{w'}(\varphi) = \True$ \\ 
for some $w'$ such that $(w, w', I_w(a)) \in R$ 


\item $I_w(\Q_\nil(\varphi)) = \begin{cases}
            				1, & \text{if } I_w(\varphi) = \True \\
            				0, & \text{if } I_w(\varphi) = \False
                         \end{cases} $

\item $I_w(\Q_{a\cons l}(\varphi)) = \sum\limits_{w' | (w, w', I_w(a)) \in R } \probab((w,w',I_w(a))) . I_{w'}(\Q_l(\varphi)) $
\end{itemize}
\end{small}
\end{definition}

In the probabilistic logic \PTL, validity and satisfaction of a formula by a model are standard non-probabilistic notions, as defined below. The logic handles probabilities explicitly in its language; not at the semantic level. 

\begin{definition}
\label{def:Satisfaction}
A formula $\varphi$ is \emph{satisfied} in a model $M \equiv (W, R, \probab, \{ D_{\tau} \}_{\tau \in T}, I)$ in a state $w$, denoted $M, w \vDash \varphi$ iff $I_w(\varphi) = \True$. A formula $\varphi$ is \emph{globally satisfied} in a model $M$, denoted $M \vDash \varphi$ iff $M, w \vDash \varphi$ for all $w \in W$. A formula $\varphi$ is \emph{valid}, denoted $\vDash \varphi$ iff $M \vDash \varphi$ for every model $M$. A set of formulas $T$ entails a formula $\varphi$, denoted $T \vDash \varphi$, iff $M \vDash \varphi$ for every model $M$ such that $M \vDash \bigwedge_{G \in T} G$.  
\end{definition}

\section{Adequacy}
\label{sec:Adequacy}

This section shows how the usual mathematical presentation of probability theory, as recalled in Definition \ref{def:ProbabilitySpace}, can be considered a special case of the probabilistic logic presented here. This is done by showing (in Theorem \ref{theorem:Adequacy}) how to translate probability spaces into models and the usual set-theoretic language for probabilistic events into \PTL's language.

\begin{definition}
\emph{Set expressions} over a set $\Omega$ are expressions freely generated from singleton subsets of $\Omega$ and operators for complementation ($\overline{\phantom{\{ \}}}$), union ($\cup$) and intersection ($\cap$).  
\end{definition}

\begin{remark}
As usual, by abuse of notation, set expressions and the sets they denote are not explicitly distinguished.
\end{remark}

\begin{example}
If $\Omega = \{ w_1, w_2 \}$ then the following are examples of set expressions: $\{w_1\}$, $\{w_2\}$, $\overline{\{w_2\}}$ (denoting the set $\{ w_1 \}$), $\{w_1\} \cup \{w_2\}$ (denoting the set $\{w_1, w_2\}$), $\{w_1\} \cap \{w_2\}$ (denoting the empty set), \ldots
\end{example}

\begin{definition}
\label{def:ProbabilitySpace}
A \emph{probability space} is a triple $(\Omega, \Sigma, Q)$ where $\Omega$ is the \emph{sample space} (whose elements are \emph{outcomes}), $\Sigma$ is a \emph{$\sigma$-algebra} on $\Omega$ (i.e. a collection of subsets of $\Omega$ (\emph{events}) closed under complementation, countable union and countable intersection) and $Q: \Sigma \imp [0,1]$ is a probability function satisfying Kolmogorov's axioms: 
\begin{enumerate}
\item $Q(E) \geq 0$, for all $E \in \Sigma$
\item $Q(\Omega) = 1$
\item For any countable collection $C$ of mutually disjoint events 
$$Q(\bigcup_{E \in C} E) = \sum_{E \in C} Q(E)$$ 
\end{enumerate}
\end{definition}

\begin{theorem}
\label{theorem:Adequacy}
For every probability space $(\Omega, \Sigma, Q)$, there is a model $M$ and a language translation function $g$ from set expressions over $\Omega$ to formulas such that $Q(E) = p$ iff $M, w \vDash \Q_a(g(E)) = p$, for some $w$ and some $a$.
\end{theorem}
\begin{proof}
Let $W$ be $\{ w \} \cup \Omega$. For each $w_k \in \Omega$, let $F_k$ be a distinct atomic proposition. Let $I$ be any interpretation such that $I_{w_i}(F_j) = \True$ iff $i = j$. Let $R$ be $\{ (w, w_k, I_w(a)) | w_k \in \Omega \}$. Let the probabilistic transition function be defined such that $\probab((w, w_k, I_w(a))) = Q(\{w_k\})$. Since $\Omega = \bigcup_k \{w_k\}$, all $\{w_k\}$ are mutually disjoint and $Q(\Omega) = 1$, the condition (from Definition \ref{def:Frame}) that 
$$
\sum_{w_k ~ | ~ (w, w_k, I_w(a)) \in R } \probab((w,w_k,I_w(a))) = 1 
$$ holds. Finally let $M$ be the model $(W,R,\probab,\{D_{\tau}\}_{\tau \in T},I)$.
The translation function $g$ is defined recursively:
\[
g(E) = \begin{cases}
           F_k, & \text{if } E = \{ w_k \} \\
           g(E') \vee g(E''), & \text{if } E = E' \cup E'' \\
           g(E') \wedge g(E''), & \text{if } E = E' \cap E'' \\
           \neg g(E'), & \text{if } E = \overline{E'}
        \end{cases}
\]
Now the fact that $Q(E) = p$ iff $M, w \vDash \Q_a(g(E)) = p$ must be proven. First notice that, by Definition \ref{def:Satisfaction}, $M, w \vDash \Q_a(g(E)) = p$ iff $I_w(\Q_a(g(E)) = p)$, and by Definition \ref{def:Model}, $I_w(\Q_a(g(E)) = p)$ iff 
$$
\sum_{w' ~ | ~ (w, w', I_w(a)) \in R } \probab((w,w',I_w(a))). I_{w'}(\Q_{\nil}(g(E))) = p
$$ 
By Definition \ref{def:Model} again and the definition of $R$, the summation above can be simplified, resulting in the following equation: 
$$
\sum_{w' | w' \in \Omega \text{ and } I_{w'}(g(E)) = \True \}} \probab((w,w',I_w(a))) = p
$$
Furthermore, unfolding the definition of $\probab$, the equation above reduces to:
$$
\sum_{w' | w' \in \Omega \text{ and } I_{w'}(g(E)) = \True \}} Q(\{w'\}) = p
$$
Therefore, it suffices to prove that the equation above holds iff $Q(E) = p$, or equivalently, that:
$$
\sum_{w' | w' \in \Omega \text{ and } I_{w'}(g(E)) = \True \}} Q(\{w'\}) = Q(E)
$$
By Kolmogorov's third axiom, $Q(E) = \sum_{w' \in E} Q(\{w'\}$. Hence, letting $X$ be the following set:
$$
\{ x | x \in \Omega \text{ and } I_x(g(E)) = \True \}
$$
A sufficient condition for the equation above to hold is that $X = E$.
This is proven below by induction on the structure of $E$:
\begin{itemize}
\item \textbf{Base case} ($E = \{w_k\}$): then $g(E_k) = F_k$ and, by definition of $I$, $I_x(F_k) = \True$ iff $x = w_k$. 

\item \textbf{Induction cases}:
  \begin{itemize}
  \item ($E = \overline{E'}$): then $g(E) = \neg g(E')$ and hence:
  $$
  X = \{ x | x \in \Omega \text{ and not } I_x(g(E')) = \True \}
  $$
  Let  
  $$
  Y = \{ x | x \in \Omega \text{ and } I_x(g(E')) = \True \}
  $$
  By induction hypothesis, $Y = E'$. Therefore, $X = \Omega \setminus Y = \Omega \setminus E' = E$.

  \item ($E = E' \cap E''$): then $g(E) = g(E') \wedge g(E'')$ and hence:
  $$
  X = \{ x | x \in \Omega \text{ and } I_x(g(E') \wedge g(E'')) = \True \}
  $$
  and so:
  $$
  X = \{ x | x \in \Omega \text{ and } I_x(g(E')) = \True \text{ and } I_x(g(E'')) = \True \}
  $$
  Let:
  $$
  Y = \{ x | x \in \Omega \text{ and } I_x(g(E')) = \True \}
  $$
  $$
  Z = \{ x | x \in \Omega \text{ and } I_x(g(E'')) = \True \}
  $$
  By induction hypothesis, $Y = E'$ and $Z = E''$. Therefore, $X = Y \cap Z = E' \cap E'' = E$.

  \item ($E = E' \cup E''$): this case is analogous to the case above.
  \end{itemize}
\end{itemize}
\end{proof}





Informally, the idea of the proof of Theorem \ref{theorem:Adequacy} is to translate a probability space into a model with a distinguished initial state and a future state for each possible outcome in the space. Any set expression (specifying an event) has a corresponding logical formula. The correspondence is as expected: union corresponds to disjunction, intersection to conjunction and complementation to negation. The translation is \emph{adequate} in the sense that the probability of an event in the space is equal to the probability of the corresponding formula in the model.  



\section{Expressiveness}
\label{sec:Expressiveness}

A corollary of Theorem \ref{theorem:Adequacy} is that the probabilistic logic \PTL is more expressive than classical probability theory, in two distinct informal senses. The first one is syntactical: whereas the usual language of classical probability theory (which relies on set expressions) can naturally express formulas containing propositional connectives such as negation, conjunction and disjunction (through the inverse of the translation function $g$ defined in the proof of the theorem), there are formulas in \PTL's language (e.g. formulas containing quantifiers or nested probability operators) which have no (natural) counterpart in the language of classical probability theory. The second one is semantical: the proof of Theorem \ref{theorem:Adequacy} shows that probability spaces correspond to models with a very simple frame; it would be inconvenient to express models with more complex frames in terms of probability spaces, because the frame structure would have to be flattened.

\subsection{Independence}
\label{sec:Independence}

Shortcomings and limitations of probability spaces for knowledge representation become apparent in situations where a sequence of independent actions is performed over time. Suppose that a fair coin is tossed twice. Representing this as a probability space requires a sample space with four outcomes $\{ h_1h_2, h_1t_2, t_1h_2, t_1t_2 \}$. Saying that, for instance, $P(\{h_1h_2\}) = P(H_1 \cap H_2) = P(H_1) P(H_2) = 0.25$ (where $H_1 = \{h_1t_2,h_1h_2\}$ and $H_2 = \{h_1h_2,t_1h_2\}$) requires the assumption of independence for the tosses. Two events $E_1$ and $E_2$ are often defined to be \emph{independent} if and only if $P(H_1 \cap H_2) = P(H_1) P(H_2)$. But this definition is epistemologically unsatisfactory. How do we actually come to know that $H_1$ and $H_2$ are independent? According to this definition, we must know $P(H_1 \cap H_2)$ in advance. But that is precisely what, in practice, we do \emph{not} know and would like to compute (based on our knowledge of $P(H_1)$ and $P(H_2)$)! We can easily get trapped in circular reasoning, trying to justify, for instance, our claim that $P(H_1 \cap H_2) = 0.25$ by saying that $H_1$ and $H_2$ are independent and then trying to justify that they are independent by saying that $P(H_1 \cap H_2) = 0.25$. Of course, we tend to escape from such cases of fallacious circular reasoning by simply assuming that the events are independent. However, the assumption is \emph{tacit}. Classical probability theory provides no way to represent knowledge of the independence and any reason that we might have for justifying the assumption of independence of the events remains at an informal level, external to the representation.

In the \PTL, on the other hand, the possibility to represent independence comes naturally and for free. For instance, when the axiom $\al x:Coin. \ldia_{t(x)}^{0.5}. H(x) \wedge \ldia^{0.5}_{t(x)}T(x)$ is assumed, it follows from the semantics of the logic that it holds in any state of the model. And since the axiom states the equal probabilities for heads and tails in a way that does not depend on anything except the action of the toss itself, it is clear that tossing a coin at a state $s$ has no effect on tossing the coin at another state $s'$. Therefore, the two tosses must be independent, and consequently it follows that:
$$
\al x:Coin. \ldia_{t(x)}^{0.5}. H(x) \wedge \ldia^{0.5}_{t(x)}T(x) \vDash \Q_{t(x)\cons t(x)\cons \nil}(H(x)\cons H(x) \cons \nil) = 0.25
$$

Also dependence can be easily represented. For example, consider a magical coin $c_m$ that behaves as a fair coin in an initial state, but when tossed in any other state always gives the opposite result of the previous toss. This may be represented by the following axioms:
\begin{itemize}
\item $@_s \ldia_{t(c_m)}^{0.5}. H(c_m) \wedge \ldia^{0.5}_{t(c_m)}T(c_m)$

\item $T(c_m) \imp \ldia_{t(c_m)}^{1}. H(c_m)$

\item $H(c_m) \imp \ldia_{t(c_m)}^{1}. T(c_m)$

\item $\lbox \neg \In(s)$ \quad  (no state is a predecessor of $s$)
\end{itemize}

The inadequacy of classical probability theory's usual definition of independence can be further illustrated in a situation where we have to randomly get an object from a bag with four objects: a black sphere, a white sphere, a black cube and a white cube. For simplicity, we assume tacitly that we put the object back in the bag after the action. This can be represented by the following axioms: 
\textbf{A1:} $S(s_b) \wedge B(s_b)$; 
\textbf{A2:} $S(s_w) \wedge W(s_w)$; 
\textbf{A3:} $C(c_b) \wedge B(c_b)$;
\textbf{A4:} $C(c_w) \wedge W(c_w)$;
\textbf{A5:} $\mathit{Bag} = s_b \cons s_w \cons c_b \cons c_w \cons \nil$; and
\textbf{A6:} $\al x \in \mathit{Bag}. \ldia_{a}^{1/|\mathit{Bag}|} G(x)$.





It then follows, by the semantics, that:
$$
\textrm{A1}, \textrm{A2}, \textrm{A3}, \textrm{A4}, \textrm{A5}, \textrm{A6} \vDash \al x \in \mathit{Bag}. \Q_a(S(x) \wedge B(x)) = \Q_a(S(x)) . \Q_a(B(x))
$$

Nevertheless, we should not be willing to conclude from this result, as classical probability theory does, that the event of getting a spherical object and the event of getting a black object are independent. It is merely coincidental that $\Q_a(S(x) \wedge B(x)) = \Q_a(S(x)) . \Q_a(B(x))$. If the bag had an additional black tetrahedral, for instance, the two sides of this equation would not be equal anymore. In the formalization above, it is evident that both events are correlated, because they consist of outcomes from a single action.

A simple formal theory $T_{\mathit{indep}}$ of (in)dependence of actions could provide the following definition for \emph{independence} of an action $a$ from an action $b$:
\begin{itemize}
\item $ \mathit{Independent}(a,b) \equiv (\al s. @_s ((\al \varphi. \al p. \Q_a(\varphi) = p \imp \lbox_b (\Q_a(\varphi) = p))) $

\end{itemize}

\noindent
It follows from the semantics that $T_{\mathit{indep}}$ entails the following \emph{shortcut} theorem:

\medskip

\noindent
\begin{scriptsize}
$$\bigwedge_{1\leq i < j \leq n} \mathit{Independent}(a_i,a_j) \imp 
\Q_{a_1\cons\ldots\cons a_n \cons \nil}(E_1\cons\ldots\cons E_n \cons \nil) = \Q_{a_1}(E_1)\ldots\Q_{a_n}(E_n)$$
\end{scriptsize}

\medskip

The notion of independence defined in $T_{\mathit{indep}}$ is non-circular. We may, from the logical specification of a system in the \PTL's language, explicitly reason about the actions of the system, conclude that some of them are mutually independent and use the general theorem above as a shortcut for computing probabilities of sequences of actions. This is arguably more satisfactory than the teleological definition of independence from classical probability theory, which depends on the very shortcut theorem that we would have liked to derive.

It is not an aim of this paper to discuss $T_{\mathit{indep}}$ or other theories of independence in detail. $T_{\mathit{indep}}$ is just a (very simple) example showing that \PTL is expressive enough to allow explicit reasoning about concepts that are very relevant in a probabilistic context.

\subsection{Disambiguation}
\label{sec:Disambiguation}

 Informal statements about probabilities are sometimes imprecise and ambiguous. Their intended meanings are not always clear. If a person $A$ tried to describe to a person $B$ the random effects of an action $a$, her description might include a sentence such as: ``the probability of $\varphi$ after $a$ is $p$''. The most straightforward and literal logical meaning for this sentence would be $\Q_a(\varphi) = p$. However, it is often the case that the meaning intended by $A$ is actually $\ldia_a^p \varphi$. $B$ must guess, from the context of the conversation and the common knowledge, which of the two alternatives is actually meant. 

A formula such as $\ldia_a^p \varphi$ provides fine-grained information about one particular state transition that is made possible by the action, whereas $\Q_a(\varphi) = p$ provides coarse-grained aggregated information about transitions to all states where $\varphi$ holds. The aggregated information is incomplete, because it doesn't say how many such states there are and it doesn't specify the transition probability to each of these states. 

The power to disambiguate is an interesting qualitative criterium to estimate the usefulness of a formal language. The formal probabilistic logical language proposed here is expressive enough to precisely disambiguate between $\ldia$ and $\Q$, which are subtly but importantly different in meaning, even though they are often expressed indistinguishably in natural language. 
It is important to note that neither $\ldia_a^p \varphi \imp \Q_a(\varphi) = p$ nor $\Q_a(\varphi) = p \imp \ldia_a^p \varphi$ is valid. Understanding the difference between $\ldia_a^p \varphi$ and $\Q_a(\varphi) = p$ is crucial for a correct use of \PTL. Furthermore, the difference in the meanings of $\ldia$ and $\Q$ is essential to a semantics for probabilities that is compatible with our intuition about probabilities. Therefore, any sufficiently rich probabilistic logic should strive to distinguish between these important notions. \PTL does so explicitly and syntactically.

\begin{remark}
In natural language dialogues, $B$ tends to cope with the ambiguity by subconsciously attempting to presuppose that $\varphi$ fully specifies a single outcome of $a$, in which case $A$ means $\ldia_a^p \varphi$. If this presupposition is incompatible with pre-existing knowledge or even with knowledge acquired later during the dialogue, the presupposition is canceled and the meaning falls back to $\Q_a(\varphi) = p$. Fully understanding the dynamics of presuppositions is an open linguistic challenge, and probabilities bring yet another dimension of complexity to this difficult problem. 
\end{remark}

\begin{example}
Consider the following statement: 
\begin{itemize}
\item ``the probability of picking number $n$ (for $1 \leq n \leq 6$) is $1/6$'' 
\end{itemize}
Upon hearing this sentence, we tend to presuppose that there are six outcomes (i.e. $\ldia_{\mathrm{pick}}^{1/6} \mathrm{Picked}(n)$). However, if we are later told that:
\begin{itemize}
\item ``the number is picked by throwing a 12-faced fair dice where each $n$ (for $1 \leq n \leq 6$) occurs in two distinct faces.'' 
\end{itemize}
we are forced to cancel our presupposition and revise our logical interpretation of the previous sentence.
\end{example}

\section{Implementation and Automation}
\label{sec:Implementation}

A preliminary implementation of \PTL in \texttt{Coq} is available in \url{https://github.com/Paradoxika/ProbLogic}. It follows the embedding methodology used in \cite{ECAI,CSR}, which is based on a higher-order and typed version of the standard translation of modal logics into predicate logic, with three important differences. Firstly, whereas in the standard translation the accessibility relation is a primitive constant, in the embedding of \PTL it is derived from the primitive notion of action. Secondly, the higher-order modal logics used in \cite{ECAI} were \emph{rigid}, while \PTL includes a flexible probability function $\Q$ (which is simulated by a flexible predicate in the implementation). Finally, in contrast to the logics from \cite{ECAI}, \PTL requires numerical reasoning. It is this last point that makes the embedding of \PTL significantly harder than previous embeddings and justifies its preliminary status. The current implementation still does not provide convenient modal tactics (as those described in \cite{CSR}) and numerical reasoning is done with \texttt{Coq}'s standard \texttt{QArith} library for rationals (instead of real-closed fields). 
%
Decidability (of the satisfiability, validity and entailment problems) is indeed, of course, hopeless for the proposed \emph{higher-order} logic. But even for logics with undecidability issues, automated theorem provers are occasionaly sufficiently efficient for practical applications \cite{ECAI}. It is also important to note that, even if arithmetical expressions (of type $\eta$) are restricted to be ground (i.e. by forbidding quantifiers of type $(\eta \imp o) \imp o$), \PTL thus restricted would still be expressive enough to formalize all the examples shown in this paper. In this restricted logic, the only automation of arithmetic needed is simplification/computation of arithmetic expressions and reduction of ground simplified arithmetic propositions to $\top$ or $\bot$. With the recent progress in SMT-solving and automated theorem proving modulo arithmetic (even with quantifiers), it is reasonable to hope that automated provers will soon be able to cope with \PTL problems. 
In the meanwhile, the current implementation in \texttt{Coq} has already proven to be sufficient for a fully interactive formalization of the Monty Hall problem, as described in the next section.


\section{The Monty Hall Problem}
\label{sec:MontyHall}

\newcommand{\doors}{D}
\newcommand{\pick}{p}
\newcommand{\hide}{h}
\newcommand{\switch}{s}
\newcommand{\notswitch}{\bar{s}}
\newcommand{\open}{o}

\PTL is used here in the formalization of vos Savant's famous \emph{Monty Hall problem} \cite{MontyHall}, whose description is reproduced below:

\medskip

\noindent
\emph{Suppose you're on a game show, and you're given the choice of three doors:
Behind one door is a car; behind the others, goats.
You pick a door, say No. 1, and the host, who knows what's behind the doors, 
opens another door, say No. 3, which has a goat. He then says to you, 
`Do you want to pick door No. 2?' 
Is it to your advantage to switch your choice?} 

\medskip

\noindent
This probabilistic puzzle is seemingly paradoxical, because people very often make mistakes when they reason \emph{informally} about the problem, as they tend to wrongly compute the probabilities. Therefore, despite its apparent simplicity, this problem is an interesting benchmark for evaluating \emph{formal} probabilistic logics. A good probabilistic logic should allow a sufficiently natural and unambiguous formal representation of the problem and should entail correct probability values. From the player's point of view, the Monty Hall problem can be formalized in \PTL by the following axioms: 
\begin{itemize}
\item \textbf{Axiom 1:} ``you're given the choice of three doors'':
$
\doors = d_1 \cons d_2 \cons d_3 \cons \nil
$

\item \textbf{Axiom 2:} ``behind one door is a car'':
$
\ex d \in \doors. C(d)
$

\item \textbf{Axiom 3:} ``behind the others, goats'':
$
\al d \in \doors. \neg C(d) \biimp G(d)
$

\item \textbf{Axiom 4:} ``you pick a door, say No. 1, and the host, who knows what's behind the doors, 
opens another door, say No. 3, which has a goat.'':
$$
\ex s_c. ((@_{s_0} \ldia_{\hide} \ldia_{\pick(d_1)} \ldia_{\open} \In(s_c)) \wedge @_{s_c} (O(d_3) \wedge G(d_3)) )
$$
\end{itemize}

A more literal reading of Axiom 4 would be that ``there exists a state (the current state), reachable from the initial state by the sequence of actions in which the host hides the car ($\hide$), the player picks the first door ($\pick(d_1)$), and the host opens a door ($\open$), where the third door is open and has a goat.''. It is fair to say that the axioms shown above capture the intended meanings of their corresponding informal natural language sentences. As desired, the axioms are reasonably similar to the corresponding sentences, although there are interesting differences worth discussing, particularly in relation to Axiom 4. Firstly, it illustrates the need for the hybrid logic operators $@$ and $\In$ in situations where it is important to declare \emph{local} conditions, which hold only in a single given state. Secondly, it shows the convenience of having a versatile approach to actions. The $\mathit{pick}$ ($p$) action, for instance, takes the picked door as an argument whereas the $\mathit{open}$ ($o$) action takes no argument. This allows us to express that, from the point of view of the player, the action of picking a door is an action of the player and he can choose which door to pick, while opening a door is an action performed by the host, with uncertain outcomes to the player. The opening of the third door is represented as a random event of the action, through the proposition $O(d_3)$. These subtle differences between Axiom 4 and its corresponding sentence in the informal description of the problem are evidence that, as expected from a formal language, \PTL offers a higher degree of precision than what we are used to in natural language.

There are many assumptions that are not explicitly mentioned in the description of the problem. But they must be formalized as well. We list below only some of them. Other axioms (e.g. stating what remains unchanged when actions are excuted) can be see in the \texttt{Coq} formalization discussed in Section \ref{sec:Implementation}.

\begin{itemize}
\item \textbf{Axiom 5:} Each door has equal probability of having the car after the $\mathit{hide}$ ($h$) action:
$
\al d \in \doors. \ldia_{\hide}^{1/|\doors|} C(d)
$
\item \textbf{Axiom 6:} The $\mathit{pick}$ ($p$) action marks the picked door:
$
\al d \in \doors. \ldia_{\pick(d)}^1 P(d)
$
\item \textbf{Axiom 7:} The host opens a door containing a goat with uniform probability among the doors that are neither picked nor contain a car:
\begin{small}
$$
\al d^c. \al d^p. (C(d^c) \imp P(d^p) \imp \al d \in ((\doors - d^c) - d^p). \ldia_{\open}^{1/|((\doors - d^c) - d^p)|} O(d) )  
$$
\end{small}
\item \textbf{Axiom 8:} When the player does the \emph{switch} ($\switch$) action, the newly picked door is different from the previously picked door and from the open door:
\begin{small}
$$
\al d^o. \al d^p. (O(d^o) \imp P(d^p) \imp \ex d. (d \neq d^o \wedge d \neq d^p \wedge \ldia_{\switch}^1 P(d) ) )  
$$
\end{small}
\item \textbf{Axiom 9:} When the player does the \emph{no switch} ($\notswitch$) action, the newly picked door is the same as the previously picked door:
$
\al d. (P(d) \imp \ldia^1_{\notswitch} P(d) ) )  
$
\item \textbf{Axiom 10:} A state is a victorious state if and only if the car is behind the picked door:
$
V \leftrightarrow (\ex d. C(d) \wedge P(d))
$
\end{itemize}

The next step is the formalization of (the intended meaning of) the question (``Do you want to pick door No. 2? Is it to your advantage to switch your choice?'') as a conjecture. However, this is significantly less straightforward than the formalization of the axioms. A naive and literal reading of the question could result in the following tentative conjecture:
$$
\Q_{\switch}(V) > \Q_{\notswitch}(V)
$$
But the formula above is only satisfied in models where the probability of victory by switching is greater than the probability of victory by not switching in \emph{all} states, whereas the question is interested in a few states only, namely those reachable by a given sequence of actions (i.e. hiding, picking, opening and re-picking).
Taking this into account, an apparently plausible alternative formalization could be:
$$
@_{s_0} \lbox_{\hide } \lbox_{\pick(d_1)} \lbox_{\open} (\Q_{\switch}(V) > \Q_{\notswitch}(V))
$$
But this is trivially false in any model $M$ that satisfies the axioms above, because the action $\mathit{hide}$ has a successor state $s_1$ (where the car was hidden behind the first door) such that:
$$
M \vDash @_{s_1} \lbox_{\pick(d_1)} \lbox_{\open} (\Q_{\switch}(V) < \Q_{\notswitch}(V))
$$
Yet another possible attempt would be to formalize the conjecture as:
$$
\forall s. \varphi(s) \imp @_s \Q_{\switch}(V) > \Q_{\notswitch}(V)
$$
where $s$ is the current state when the question is asked and $\varphi(s)$ is a formula specifying whether $s$ is a posible current state (i.e. consistent with the player's observations). However, for a similar reason, this formula is also false in any model $M$ that satisfies the axioms: there is a possible current state $s^*$, where $I_{s^*}(\Q_{\switch}(V)) = 0$ and $I_{s^*}(\Q_{\notswitch}(V)) = 1$. In fact, it is easy to see that, in any possible current state $s$, $I_{s^*}(\Q_{\notswitch}(V))$ and $I_{s^*}(\Q_{\switch}(V))$ are always either $0$ and $1$, because the action of switching has always only one possible outcome.

As evidenced by the failed conjectures above, there is a structural gap between the natural language question and the correct formalization of its intended meaning, and therein lies a potential reason (though probably not the only one) why people tend to have difficulties to reason about the Monty Hall problem. As it is posed, the question induces the player to think in terms of probabilistic outcomes of the action of switching or not switching in the current state. In contrast, the correct thinking requires the player to hypothetically backtrack to the initial state and formulate the conjecture as follows:

\begin{itemize}
\item \textbf{Conjecture: }
$
@_{s_0}
(\Q_{\hide \cons \pick(d_1) \cons \open \cons \switch \cons \nil}(V) 
> 
\Q_{\hide \cons \pick(d_1) \cons \open \cons \notswitch \cons \nil}(V))
$
\end{itemize}

\noindent
In any model satisfying the axioms (including the omitted axioms), $I_{s_0}(\Q_{\hide \cons \pick(d_1) \cons \open \cons \switch \cons \nil}(V)) = 2/3$ and $I_{s_0}(\Q_{\hide \cons \pick(d_1) \cons \open \cons \notswitch \cons \nil}(V)) = 1/3$. Therefore, the conjecture is a theorem\footnote{An interactive proof of this theorem using the embedding of \PTL in \texttt{Coq} is freely available in the online repository of the implementation discussed in Section \ref{sec:Implementation}. For the sake of simplicity, this formalization of the Monty Hall problem does not concern itself with specifying in which states each action is allowed or disallowed. But this could also be done.}.






\section{Related Work}
\label{sec:RelatedWork}

Many probabilistic logics are surveyed in \cite{SEP}. Among those logics, most depart from classical logic by adopting a probabilistic notion of validity and entailment. \PTL, on the other hand, remains strictly classical in this respect. The probabilistic modal logics described in Sections 4.1 and 4.2 of \cite{SEP} are probably the most similar to \PTL. However, they are propositional, lack the probabilistic diamond operator, and are atemporal.

Probabilistic logics that incorporate time include \textbf{PCTL} \cite{PCTLOriginal,PCTL}, which extends \textbf{CTL} by replacing the existential and universal path quantifiers by a probabilistic operator. \textbf{PCTL} is an excellent logic for \emph{model checking} Markov chains. However, its lack of a probabilistic diamond operator makes it susceptible to the issues discussed in Section \ref{sec:Disambiguation}, thereby limiting its use beyond model checking. They also lack an explicit handling of actions, which is necessary for a convenient formalization of the Monty Hall problem and other examples discussed here. On the other hand, \textbf{PCTL}'s temporal modalities (which include, for instance, the \emph{until} operator) are more sophisticated than \PTL's temporal modalities (which can only make statements about the \emph{next} moment in time). \PTL's parsimony is intentional: it includes only the minimal set of temporal modalities needed to capture the desired notion of probability. Nevertheless, in practical applications where other temporal modalities are needed, they could be easily added to \PTL as well.  


\section{Conclusion and Future Work}
\label{sec:Conclusion}

The large number of available probabilistic logics indicates that conciliating logic and probability is a non-trivial task. The expressive probabilistic temporal logic \PTL described here provides a novel alternative approach, based on the simple intuition that the notion of probability can only be fully grasped in combination with the notions of action and time. The complex interaction of time, action and probability naturally leads to a modal and higher-order logic. \PTL is adequate with respect to classical probability theory, of which it can be considered an extension (as shown in Section \ref{sec:Adequacy}, where a correspondence between events and formulas has been established in detail). \PTL's convenient expressive power allowed a natural formalization of the famous Monty Hall problem. One of the main insights in the development of \PTL came with the discovery of the need for both a higher-order probability function and a probabilistic diamond operator, as discussed in Section \ref{sec:Disambiguation}. Besides the higher order, the satisfaction of this need is a distinguishing feature of \PTL.

In the near future, the implementation of \PTL in \texttt{Coq} needs to be made more user-friendly, through the implementation of tactics that automate and hide technical details for users. 
On the philosophical side, it would be interesting to extend \PTL with past temporal modalities, since we often need to reason about actions that have happened in the past but whose outcomes we have not yet observed, and to define conditional probabilities, in order to explore the question about the relationship between probabilities of conditionals (e.g. $P(A\imp B)$) and conditional probabilities (e.g. $P(B|A)$) \cite{HajekProbabilities} from \PTL's perspective.









\bibliographystyle{aiml16}
\bibliography{bibliography}

\end{document}